\documentclass[aps,prl,reprint]{revtex4-1}
\usepackage{amsmath}
\usepackage{graphicx}
\usepackage{units}  
\usepackage{xspace}
\usepackage{subfigure}
\usepackage{hyperref}

\newcommand{\mb}{\mathbf}
\newcommand{\tb}{\textbf}
\newcommand{\beq}{\begin{equation}}
\newcommand{\eeq}{\end{equation}}
\newcommand{\bea}{\begin{eqnarray}}
\newcommand{\eea}{\end{eqnarray}}

\usepackage[usenames]{color}

\begin{document}

\bibliographystyle{apsrev}
 
\title{On magic angles and band flattening in twisted bilayer graphene}

\author{Hridis K. Pal}
   \email{hpal@uh.edu}
\affiliation{Department of Physics, University of Houston, Houston, TX 77204, USA}

\begin{abstract} 
When two graphene layers are rotated from AA or AB configuration by a small angle, the band structure changes dramatically. Numerical calculations have shown that, at certain discrete angles called magic angles, the low energy bands become flat leading to localization of electrons. The origin of this strange behavior, however, is not well understood. Here, I propose a theory that offers an understanding of the phenomenon, focusing on the first magic angle. It is shown that coupling between the layers, in addition to renormalizing the Dirac velocity, introduces higher order momentum terms in the energy dispersion that are not all of the same sign. Partial cancellation among these terms leads to the flatness of the low energy bands. Also, while there is modulation of electron density in real space, there is no localization---the modulation arises due to the superposition of plane wave states with different momenta in the two layers. In addition, it is conjectured that there is an underlying geometric reason for the appearance of more than one magic angle which can be exploited to predict higher magic angles approximately without computing the band structure. 
\end{abstract}

\pacs{}
\maketitle

In recent years bilayer graphene with arbitrary angles of rotation between the layers leading to large moir{\'e} superlattices have attracted considerable attention \cite{melejphys,nori}. Such a system, dubbed twisted bilayer graphene (TBG), presents a plethora of fascinating electronic properties \cite{meleprb,kim,palprb,palarxiv,markus,roy,palarxivnat} not present in the more commonly studied AA- and AB-stacked bilayers. It has been shown that when the layers are rotated away from AA or AB configuration by a small angle, the electron velocity reduces, the effect becoming more pronounced with decreasing angle \cite{lopes,luican}. Finally, at some angle the velocity goes to zero altogether, leading to flat bands and electron localization \cite{morell,laissardiere,laissardiere2}. As the angle is decreased further, the velocity becomes nonzero but returns to zero at a smaller angle. With continuously decreasing angle, the velocity vanishes repeatedly at certain discrete angles, commonly called magic angles \cite{bistritzer}.

The problem of magic angles is challenging because it is nonperturbative. The low energy physics of TBG at small angles of rotation is described by two energy scales: interlayer coupling $\gamma$ (defined as the first Fourier component of the spatially varying interlayer coupling function) and the difference between the energies of a given momentum state in the two layers, $v\delta K$, where $v$ is the single layer velocity and  $\delta K$ is the momentum difference between the Dirac points in the two layers in a given valley ($\hbar=1$). In the regime $\gamma/v\delta K<1$, the problem can be solved perturbatively \cite{lopes}. Unfortunately, magic angles and their associated phenomena arise in the regime  $\gamma/v\delta K>1$, where perturbation theory fails \cite{bistritzer}. As a result, the phenomenon is accessible only via large scale numerical calculations, without a clear understanding of the underlying causes. 
Despite some attempts in the past \cite{nonabelian,lopes2}, questions remain: Why do flat bands arise only at magic angles? At these angles, electrons are found to localize at the AA-like regions in the superlattice depopulating AB/BA-like regions \cite{laissardiere,nonabelian}---why does this specific pattern emerge? Are these angles a fortuitous occurrence or guided by some symmetry? While these questions are important in their own right, they have become even more so in light of recent experiments where flat bands at the first magic angle have been observed experimentally \cite{nonabelianexp,mott,supercon}, along with the consequent effects of strong correlation leading to a Mott-like insulating phase \cite{mott} and a superconducting phase \cite{supercon}. An understanding of the single particle physics is indispensable for any meaningful understanding of these reported correlated behaviors.

The goal of this paper is to provide answers to the questions raised above. Focusing on the first magic angle, it is shown that coupling between the layers, in addition to renormalizing the Dirac velocity, introduces higher order momentum terms in the energy dispersion that are not all of the same sign. A vanishing velocity at the Dirac point is not sufficient to make the band nondispersive away from the Dirac point, near cancellation of these higher order terms is responsible for the phenomenon. Also, the flat band does not lead to localization of electrons in real space. Instead, the electron density modulates on the scale of the moir{\'e} supercell due to the superposition of plane wave states with different momenta in the two layers. In addition, it is conjectured that there is an underlying geometric reason for the appearance of more than one magic angle. This can help predict higher magic angles approximately without computing the band structure.

\begin{figure}
\includegraphics[scale=0.30,trim={0 1cm 7cm 0},clip]{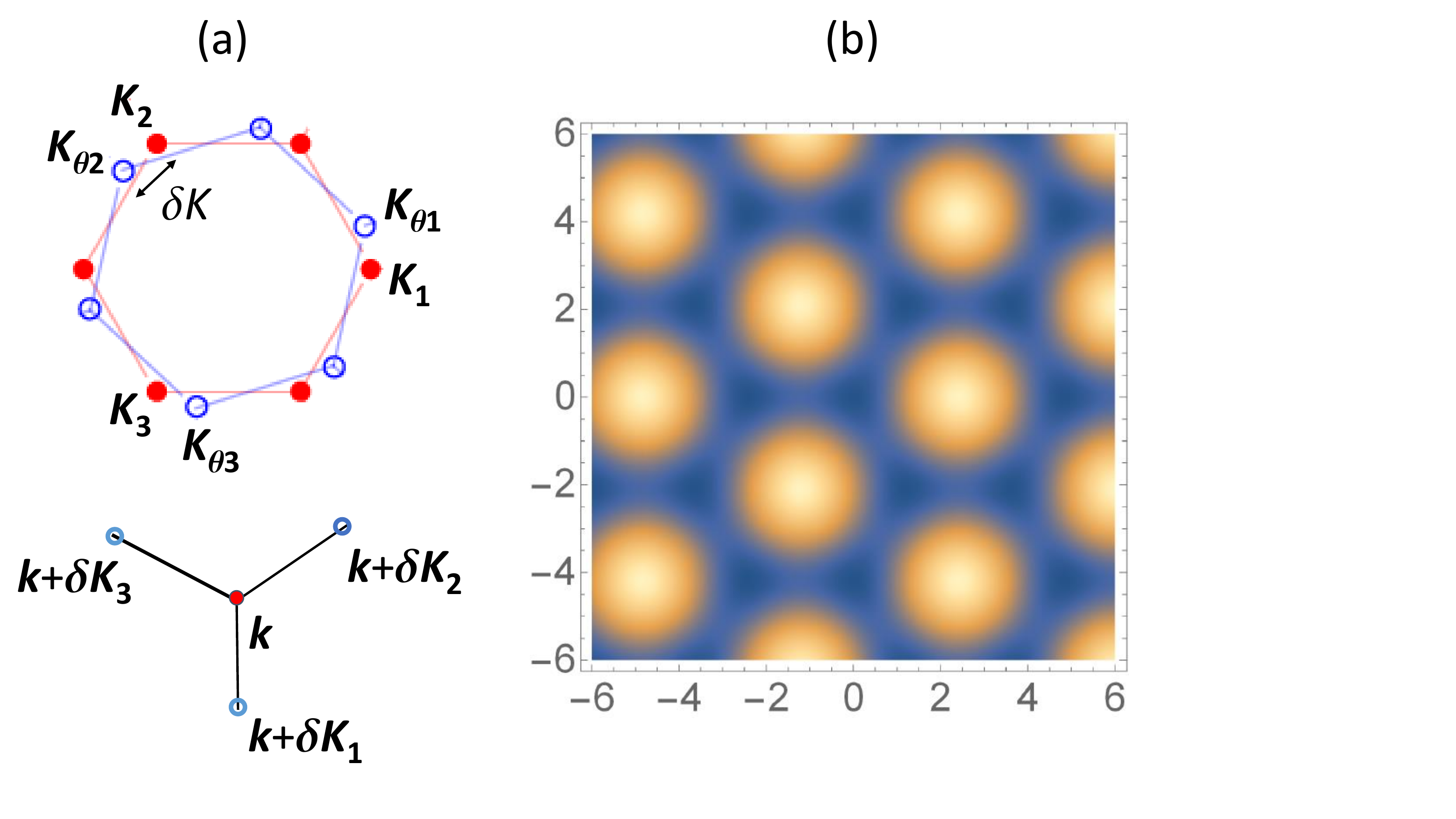}
\caption{(a) Rotation between two graphene layers leads to a rotation in reciprocal space as well (\emph{top}). For small rotations, in the simplest approximation, interlayer coupling leads to a state near the Dirac point (in a given valley) in one layer couple with three equivalent states near the Dirac point in the other layer (\emph{bottom}). (b) Spatial modulation of electron density at the first magic angle according to Eq.~(\ref{density}) (light denotes higher density). The density is maximum at AA regions and minimum at AB/BA regions with a periodicity of $4\pi/3\delta K$.}
\label{localpattern}
\end{figure}

Consider two graphene layers rotated from AB configuration by some small angle $\theta$. A rotation in real space leads to a rotation in the reciprocal space as well, as shown in Fig.~\ref{localpattern}(a). For low energy physics, one can consider states only in the vicinity of the respective Dirac points of two layers in a given valley. Interlayer coupling leads to mixing of these states. Because of the offset between the Dirac points, momentum conservation leads to coupling of states with different momenta (measured from the respective Dirac points) in the two layers. As seen in Fig.~\ref{localpattern}, a state with momentum $\mb{k}$ in one layer gets coupled to states with momenta $\mb{k}+\mb{\delta K}_{\alpha}$ in the other, where $\mb{\delta K}=\mb{K_{\theta}}-\mb{K}$, $\mb{K}$ and $\mb{K_{\theta}}$ denote the position of the unrotated and rotated Dirac points, respectively, and $\alpha=1,\cdots,3$ denotes three different orientations of $\mb{\delta K}$ differing from each other by an angle of $2\pi/3$.  When the coupling is weak, one can consider the mixing of only these states and write down a Hamiltonian following Ref.~\cite{bistritzer}:
\beq
H_{\mb{k}}=
\begin{pmatrix}
h_{\mb{k}}&T_1&T_2&T_3\\
T_1^{\dagger}&h_{\mb{k}+\mb{\delta K}_1}&0&0\\
T_2^{\dagger}&0&h_{\mb{k}+\mb{\delta K}_2}&0\\
T_3^{\dagger}&0&0&h_{\mb{k}+\mb{\delta K}_3}
\end{pmatrix},
\label{hamtrunc}
\eeq
\beq
h_{\mb{q}}=-vq
\begin{pmatrix}
0&e^{i \varphi_{\mb{q}}}\\
e^{-i \varphi_{\mb{q}}}&0
\end{pmatrix},
T_{\alpha}=\frac{\gamma}{3}
\begin{pmatrix}
e^{-i \frac{2(\alpha-1)\pi}{3}}&1\\
e^{i\frac{2(\alpha-1)\pi}{3}}&e^{-i\frac{2(\alpha-1)\pi}{3}}
\end{pmatrix},
\label{coupmat}
\eeq
where $\varphi_{\mb{q}}$ denotes the azimuthal angle. [The coupling parameter $\gamma$ is defined to be three times that used in Ref.~\cite{bistritzer}, hence the factor of $1/3$.] Solving perturbatively to linear order in momentum around the Dirac point, it was shown---first in Ref.~\cite{lopes} and later in Ref.~\cite{bistritzer}---that this leads to the preservation of the Dirac cone, albeit with a renormalized Dirac velocity. At magic angles, the Dirac velocity goes to zero.

A vanishing Dirac velocity implies that there is no linear term in momentum in the energy dispersion near the band touching point. However, it does not automatically imply a flat band: for example, although linear bands are absent in AB bilayer graphene, the bands are not flat but quadratic. Yet, numerical calculations show that the low energy bands in TBG at the magic angle, while not exactly flat, are almost dispersionless in a large area of the Brillouin zone away from the Dirac point. To unravel its origin, I consider the Hamiltonian in Eq.~(\ref{hamtrunc}) but adopt a different approach compared to Refs.~\cite{lopes} and \cite{bistritzer}: I include momentum exactly and treat the coupling perturbatively. Since the Hamiltonian is correct only to $\mathcal{O}(\gamma^2)$ to begin with, including the coupling perturbatively to the same order is justified and does not introduce any additional error. 
Without coupling, the energies are $\varepsilon^{0\pm}_{\mb{k}}=\pm v|\mb{k}|$ and $\varepsilon^{0\pm}_{\mb{k}+\mb{\delta K}_{\alpha}}=\pm v|\mb{k}+\mb{\delta K}_{\alpha}|$, with states $\psi^{0\pm}_\mb{k}=\phi^{0\pm}_{\mb{k}}e^{i(\mb{K}+\mb{k})\cdot\mb{r}}$ and $\psi^{0\pm}_{\mb{k}+\mb{\delta K}_{\alpha}}=\phi^{0\pm}_{\mb{k}+\mb{\delta K}_{\alpha}}e^{i(\mb{K}+\mb{\delta K}_{\alpha}+\mb{k})\cdot\mb{r}}$, $\alpha=1,\cdots,3$. Here, $\phi^{0\pm}_{\mb{q}}=\frac{1}{\sqrt{2}}\{\pm e^{i\varphi_{\mb{q}}},1\}$ is the Dirac spinor.  Coupling mixes these states. Focusing on one of the bands near the Dirac point, say `-', after coupling I find
\begin{widetext}
\begin{eqnarray}
&&\varepsilon^{\gamma-}_{\mb{k}}=-vk \left[1-\frac{\gamma^2}{v^2\delta K^2}\left\{\frac{1+\frac{1}{3}\frac{k}{\delta K}\mathrm{sin}[3\varphi_{\mb{k}}]-\frac{4}{3}\frac{k^2}{\delta K^2}(1+\mathrm{cos}[3\varphi_{\mb{k}}])}{1-3\frac{k^2}{\delta K^2}+2 \frac{k^3}{\delta K^3}\mathrm{sin}[3\varphi_{\mb{k}}]}\right\}\right]\label{en},\\
&&\psi^{\gamma-}_\mb{k}=\sum_{\pm}c^{\pm}\psi^{0\pm}_\mb{k}+\sum_{\pm,\alpha}c^{\pm}_{\alpha}\psi^{0\pm}_{\mb{k}+\mb{\delta K}_{\alpha}}=u_{\mb{k}}\left(\frac{\gamma}{v\delta K},\mb{\delta K}_{\alpha},\mb{r}\right)e^{i\mb{k}\cdot\mb{r}}\label{psi},\\
&&\mathrm{where}\quad u_{\mb{k}}\left(\frac{\gamma}{v\delta K},\mb{\delta K}_{\alpha},\mb{r}\right)=\sum_{\pm}c_{\mb{k}}^{\pm}\phi^{0\pm}_{\mb{k}}e^{i\mb{K}\cdot\mb{r}}+\sum_{\pm,\alpha}c_{\mb{k}\alpha}^{\pm}\phi^{0\pm}_{\mb{k}+\mb{\delta K}_{\alpha}}e^{i(\mb{K}+\mb{\delta K}_{\alpha})\cdot\mb{r}}.\label{u}
\end{eqnarray}
\end{widetext}

Consider first the expression for energy in Eq.~(\ref{en}). The linear term in momentum vanishes when $\gamma/v\delta K=1$, consistent with the results of Refs.~\cite{lopes} and \cite{bistritzer}. As anticipated, higher order terms in momentum do appear; however, surprisingly they are not all of the same sign. Therefore, as one moves away from the Dirac point, partial cancellation among these higher order terms restrict the band from becoming truly dispersive. Note that this partial cancellation is direction dependent: expanding the expression in $k$ leads to the terms alternating in sign in certain directions while in other directions all terms have the same sign. Therefore, the band is not flat to the same degree in all directions. 

The above calculation is perturbative in $\gamma$. Because $\gamma/v\delta K\sim\mathcal{O}(1)$ at the magic angle, higher order corrections in $\gamma$ are not negligible. In fact, they are important: these higher order corrections are required to redress the spurious divergence in Eq.~(\ref{en}) at certain values of $k$. A general expression for the exact energy spectrum can be written as $\varepsilon^{\gamma}_{\mb{k}}=\sum_{i=1}^{\infty}f_i(\gamma^2/v^2\delta K^2)k^i$, where $f_i$ is some function. In principle, there exists the possibility that at the magic angle $f_i\rightarrow 0$ for all $i$, leading automatically to the flat band without requiring partial cancellations between different momentum terms. This, however, can be ruled out. Assuming that $f_i$ is an analytic function of $\gamma$, while $f_{i=1}=1-\sum_{n=1}^{\infty}a_{1n}\left(\frac{\gamma^2}{v^2\delta K^2}\right)^n$, $f_{i>1}=\left(\frac{\gamma^2}{v^2\delta K^2}\right)\left[1-\sum_{n=2}^{\infty}a_{in}\left(\frac{\gamma^2}{v^2\delta K^2}\right)^n\right]$. Thus, while $f_{i=1}$ goes to zero due to the leading order effect in $\gamma$, $f_{i>1}$ can go to zero only due to next to leading order effect in $\gamma$. Therefore, $f_i$ cannot go to zero simultaneously for both $i=1$ and $i>1$. This is further evidenced by the fact that, if this were the case, the band would be flat at all $k$, which is not seen in numerics---far away from the Dirac point, the bands starts to become dispersive \cite{bistritzer}. This argument, however, does not preclude the possibility that at the magic angle, although $f_{i>1}$ is not zero, it can still be small, contributing further to the flatness. Indeed, since $f_{i=1}$ has more than one zeros (magic angles), it is guaranteed to be a nonmonotonic function of $\gamma$. Presumably, $f_{i>1}$ also shares the same behavior, and can decrease with increase in $\gamma$ in certain windows of $\gamma$. In all likelihood then the flat band arises due to a combination of two effects: partial cancellation of higher order momentum contributions and each such contribution with reduced strength, the former playing the dominant role.

Next, I turn to the wavefunction in Eq.~(\ref{psi}). Like the energy spectrum, an expression for the wavefunction can be written for any $k$ correct up to $\mathcal{O}(\gamma^2)$. But it is extremely cumbersome. Instead, I focus on the wavefunction at $k=0$ where the expression in Eq.~(\ref{psi}) simplifies greatly: $c_{\mb{0}}^-=1-\frac{\gamma^2}{3v^2\delta K^2}$, $c_{\mb{0}}^+=0$, 
$c_{\mb{0}\alpha}^-=\frac{\gamma}{3\sqrt{2}v\delta K}e^{i\pi/4}(e^{i\phi_{\mb{k}}}+e^{i2(\alpha-1)\pi/3})$, and $c_{\mb{0}\alpha}^+=-\frac{\gamma}{3\sqrt{2}v\delta K}e^{-i\pi/4}(e^{i\phi_{\mb{k}}}+e^{i2(\alpha-1)\pi/3})$. The density of electrons is given by $\lvert\psi_{\mb{0}}^{\gamma -}\rvert^2=\lvert u\left(\frac{\gamma}{v\delta K},\mb{\delta K}_{\alpha},\mb{r}\right)\rvert^2$. Averaging over all possible directions $\varphi_{\mb{k}}$ at $k=0$, I have
\begin{widetext}
\beq
\frac{|\psi_{\mb{0}}^{\gamma -}(\mb{r})|^2}{|\psi_{\mb{0}}^{\gamma -}(0)|^2}=1+\frac{1}{6}\frac{\gamma^2}{v^2\delta K^2}\left[1-\frac{1}{6}\sum_{\alpha\ne\alpha'}\left\{\mathrm{cos}[(\mb{\delta K}_{\alpha}-\mb{\delta K}_{\alpha'})\cdot\mb{r}]-\sqrt{3}\lvert\mathrm{sin}[(\mb{\delta K}_{\alpha}-\mb{\delta K}_{\alpha'})\cdot\mb{r})]\rvert\right\}\right].
\label{density}
\eeq 
\end{widetext}
When $\gamma=0$, the density is position independent, as expected in an isolated graphene layer. As the coupling is turned on, the density becomes spatially modulated. Substituting $\gamma/v\delta K=1$ at the magic angle, the electron density is found to modulate between maxima and minima following a triangular pattern, with the minimum density $2/3$ times the maximum density. The resulting pattern is shown in  Fig.~\ref{localpattern}(b). The maxima occur at the AA regions and minima at the AB/BA regions. It is consistent with previous numerical calculations \cite{laissardiere,nonabelian}. The spatial modulation arises due to superposition of plane waves with different wavelengths in Eq.~(\ref{u}). Interestingly, the sublattice degrees of freedom only changes the results quantitatively, but does not play the deciding role in the emergence of the pattern. An important observation is that even though there is spatial modulation, there is no localization. At the first magic angle, since $\gamma/v\delta K=1$, the only length scale left in Eq.~(\ref{u}) is $\delta K$, i.e., the spatial modulation happens over the superlattice length scale; thus, there is no true localization. This statement remains valid even when higher order corrections in $\gamma$ are considered. These corrections only change the condition for the first magic angle to $\gamma/v\delta K=\mathcal{O}(1)$ but does not introduce any new length scale; therefore, they cannot lead to true localization.  In the past the appearance of such a pattern has been variously attributed to (quasi)localization due to effective quantum well potentials at AA regions \cite{nonabelian} and due to momentum mismatch between states in the AA and AB regions at zero energy \cite{lopes2}. Here, it is shown that the origin is purely geometric.

\begin{figure*}
\centering
\includegraphics[scale=0.5,trim={0 8cm 0 0},clip]{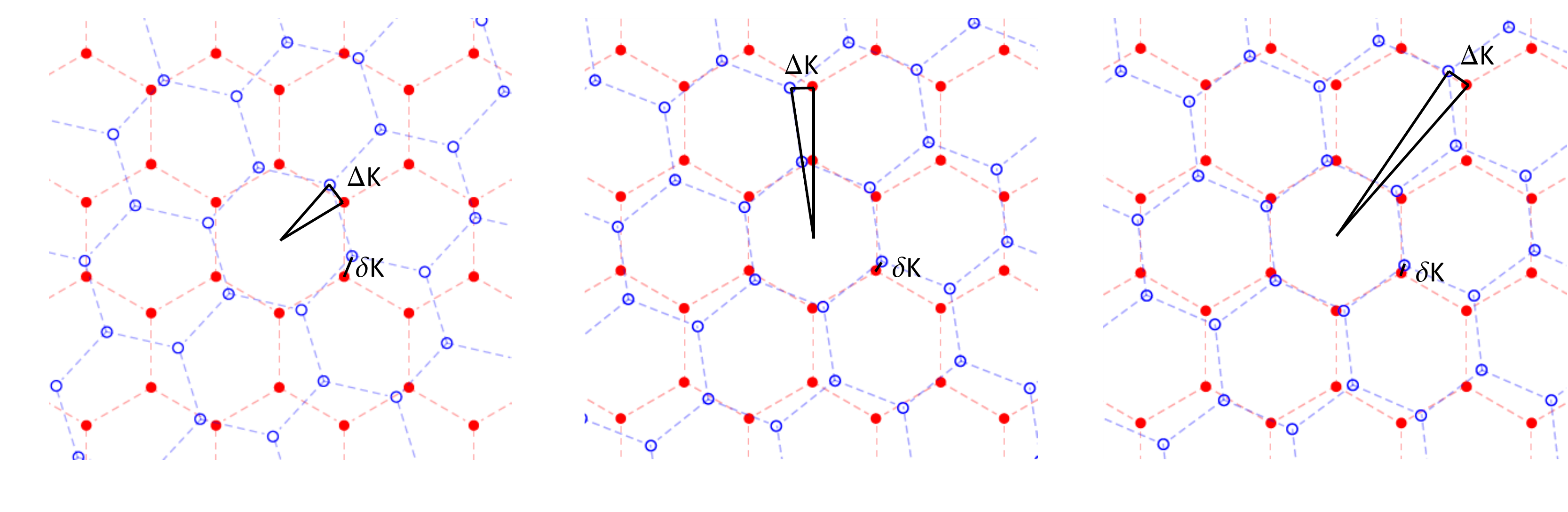}
\includegraphics[width=0.80\textwidth]{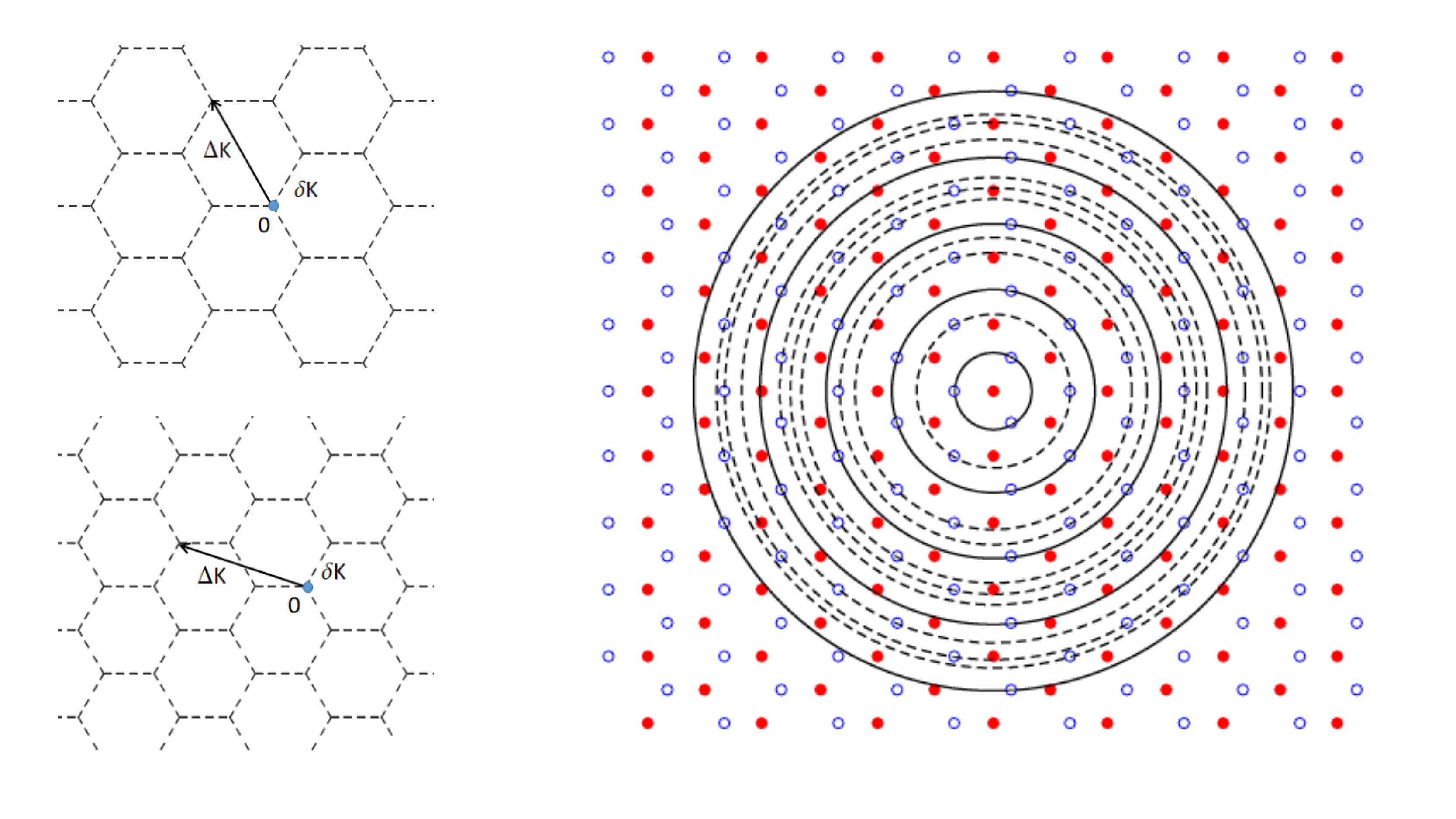}
\caption{\emph{Top}: The unrotated (red, solid circles) and rotated (blue, open circles)  layers  in momentum space. The system is resonant when the difference between the Dirac points  $\Delta K$ satisfies the condition $\gamma=v\Delta K$. While this happens once in the first Brillouin zone, in the extended Brillouin zone, the condition is satisfied repeatedly as the angle is decreased. \emph{Bottom}: Hopping between the rotated and unrotated layers sets up a hexagonal lattice. The resonance condition corresponds to $\mb{\Delta K}$ connecting the origin to a lattice point in this space. As the angle decreases, the lattice structure shrinks but $\mb{\Delta K}$ merely rotates. Shown left, there are two groups of lattice points which are three hops away from the center. These can be reached by $\mb{\Delta K}$ at two different angles. The two figures correspond to the second and third figure in the top panel. Therefore, they together contribute to the second magic angle. Shown right, this is a generic feature of all hoppings more than one: the family of dashed circles ending with a solid circle all correspond to the same number of hops from the center. }
\label{geometry}
\end{figure*}

Since there is no true localization, the low energy physics at or near magic angles cannot be described by an effective tight-binding model with lattice points at the AA sites. This is particularly important in the context of studying effects of correlation, an area that has witnessed an explosion in activity recently due to the experimental discovery of interaction driven phases in TBG at the first magic angle \cite{mott,supercon}. Clearly, writing down a Hubbard model on a triangular lattice, inspired by the triangular pattern in Fig.~\ref{localpattern}, will be ineffective. To follow this direction, one will need to first construct appropriate Wannier states---see, for example, Ref.~\cite{senthil}. Alternatively, staying within the long wavelength description, Eqs.~(\ref{en})--(\ref{u}) can serve as a good starting point to include electronic correlations. Even though they are perturbative in $\gamma$ and lack quantitative accuracy, they contain all the essential ingredients that define the single particle physics at the first magic angle.

The discussion so far has focused entirely on the first magic angle. Numerical calculations have demonstrated that, as the angle is reduced, the Dirac velocity vanishes at other angles as well \cite{bistritzer}. It is natural to ask whether this is purely by accident---some sort of `mathematical conspiracy'---or there exists a deeper reason why the velocity vanishes repeatedly. Since the problem is analytically intractable, I take recourse to physical arguments and put forward a conjecture regarding the origin of these other angles. 

Observe that once coupling is introduced, the bilayer is described by a common eigenstate, implying that a given momentum eigenstate belongs to both layers. The states describing each layer at the same energy do not have the same momentum. Alternatively, states in each layer described by the same momentum are not at the same energy. In TBG with a large angle of rotation, this difference in the energy is large---much larger than what the coupling can provide. Thus, the two layers do not couple well and electrons in each layer continue to maintain their original single layer like behavior. This, however, changes as  the rotation angle is decreased. In particular, when the interlayer coupling provides exactly the right amount of energy that compensates for the difference in energy between states in the two layers sharing the same momentum, the system can be thought to be in resonance and the process is highly favored. This occurs when $\gamma=v\Delta K$, where $\Delta K=|\mb{K}_{\theta}-\mb{K}|$ (equivalent to $\delta K$ in the first Brillouin zone, see Fig.~\ref{geometry}). This is when the first magic angle appears and the Dirac velocity goes to zero. Writing $\Delta K=2 K \mathrm{sin}\theta^{\mathrm{m}}_{1}/2\approx K\theta^{\mathrm{m}}_{1}$, one has $\theta^{\mathrm{m}}_{1}=\gamma/vK$. As the angle is changed, the system goes off resonance, and the Dirac velocity becomes nonzero.

For a magic angle to reappear, the system needs to satisfy the resonant condition once again. At first glance this seems to be impossible. However, going to the extended Brillouin zone, as seen in the top panel in Fig.~\ref{geometry}, at some angle the system does become resonant once again at the next nearest $K$ point from the origin, i.e., at angle $\theta^{\mathrm{m}}_2$, the system will satisfy $\gamma=v\Delta K$ (meanwhile, $\delta K$ in the first Brillouin zone has decreased). The process repeats periodically as the angle is decreased more and more. This argument implies that the $i-$th magic angle should be given by $\theta^{\mathrm{m}}_{i}=\frac{\gamma}{v K_i}$, 
where $K_i$ is the distance of the $i-$th nearest Dirac point from the origin in $k-$space.

\begin{table}
\caption{\label{tab1} Comparison of magic angles $\theta^{\mathrm{m}}_i$ predicted by Eq.~(\ref{magic1}) with those obtained by numerical calculations (Ref.~\cite{bistritzer}). Parameters used in computing Eq.~(\ref{magic1}) are same as in Ref.~\cite{bistritzer}: $\gamma=330$meV, $v=10^6$m/s, and $K=4\pi/3\sqrt{3}a$ with $a=1.42\AA$.}
\renewcommand\arraystretch{1.1}
\renewcommand\tabcolsep{9pt}
\centering
\begin{tabular}{ccc}
No.&$\theta^{\mathrm{m}}_i$&$\theta^{\mathrm{m}}_i$\\
$i$&[Numerics---Ref.~\cite{bistritzer}]&[This work---Eq.~(\ref{magic1})] \\
\hline\noalign{\smallskip}
1&$1.05^{\circ}$&$1.686^{\circ}$\\
2&$0.50^{\circ}$&$0.740^{\circ}$\\
3&$0.35^{\circ}$&$0.425^{\circ}$\\ 
4&$0.24^{\circ}$&$0.309^{\circ}$\\
5&$0.20^{\circ}$&$0.237^{\circ}$\\
\end{tabular}
\end{table}

This simple picture, however, overlooks an important aspect. To elucidate, I refer to the bottom panel in Fig.~\ref{geometry}. This is same as Fig.~\ref{localpattern}(a) illustrating the hopping from a state in one layer to states in another, except that processes with more than one hopping are included now. This means lattice points farther away from the origin are reached. Because the lattice in the hopping space (bottom panel in Fig.~\ref{geometry}) and the original $k-$space (top panel in Fig.~\ref{geometry}) are both hexagonal, when the condition $\gamma=v\Delta K$ is satisfied in $k-$space, $\mb{\Delta K}$ as a vector will connect the origin to a lattice point in the hopping space. The case for first resonance is trivial: $\Delta K=\delta K$. Notice, however, the second and third instances of resonance in $k-$space both correspond to three hoppings away from center in the hopping space. Thus, these do not count for two different physical instances of resonance. Instead, they will compete and the angle at which resonance is satisfied is somewhere between the angles which satisfy the individual scenarios separately. As seen in Fig.~\ref{geometry}, this is a generic feature for all hoppings more than one. With this in mind, a general formula to obtain the $i-$th magic angle can be written as 
\beq
\theta^{\mathrm{m}}_{i}=\frac{\gamma}{v} \left\langle\frac{1}{K_i}\right\rangle.
\label{magic1} 
\eeq
Here, $\langle\frac{1}{K_i}\rangle=\frac{1}{J}\sum_j\frac{1}{K_{ij}}$ is the average $\frac{1}{K}$ in the $k-$space for the $i$-th magic angle obtained by considering all the radii corresponding to the same number of hoppings in the hopping space, and $J$ is the total number of such possibilities \cite{comment}. In Table.~\ref{tab1}, the magic angles predicted by Eq.~(\ref{magic1}) are compared with those found by numerical band structure calculations. It is seen that Eq.~(\ref{magic1}) indeed approximates the magic angles.

In the discussion so far, I have ignored all hoppings at other lattice points which do not satisfy the resonance condition. They will provide corrections to Eq.~(\ref{magic1}). The difference between the predicted and numerical values stems from such corrections. These corrections are expected to be $\sim\mathcal{O}(\delta K/\Delta K)$ which should decrease at higher magic angles. This is indeed observed in Table~\ref{tab1}. Thus, Eq.~(\ref{magic1}) is conjectured to serve as a good approximation of magic angles, especially at higher magic angles where numerical calculations are prohibitively expensive.

\begin{acknowledgments}
This work was supported by financial aid from the Department of Physics and the College of Natural Science and Mathematics at the University of Houston.
\end{acknowledgments}


\begin{thebibliography}{99}
\bibitem{comment} The simple averaging adopted here is ad hoc. In reality, it could be a weighted average. However, at higher magic angles, this will lead to small corrections, much smaller than what is left out by discarding other hopping processes not at resonance.
\end{thebibliography}
\end{document}